# Graphs (Networks) with Golden Spectral Ratio

Ernesto Estrada[*]

*Complex Systems Research Group*, X-Rays Unit, RIAIDT, Edificio CACTUS, University of Santiago de Compostela, Santiago de Compostela 15782, Spain.

---

[*] Fax: 34 981 547 077. E-mail: estrada66@yahoo.com

**Abstract**

We propose two new spectral measures for graphs and networks which characterize the ratios between the width of the "bulk" part of the spectrum and the spectral gap, as well as the ratio between spectral length and the width of the "bulk" part of the spectrum. Using these definitions we introduce the concept of golden spectral graphs (GSG), which are graphs for which both spectral ratios are identical to the *golden ratio*, $\varphi = (1+\sqrt{5})/2$. Then, we prove several analytic results to finding the smallest GSG as well as to build families of GSGs. We also prove some non-existence results for certain classes of graphs. We explore by computer several classes of graphs and found some *almost* GSGs. Two networks representing real-world systems were also found to have spectral ratios very close to $\varphi$. We have shown in this work that GSG display good expansion properties, many of them are Ramanujan graphs and also are expected to have very good synchronizability. In closing golden spectral graphs are optimal networks from a topological and dynamical point of view.

# 1. Introduction

Graphs (networks) play an increasing role in our current understanding of the physical universe at different size scales. These mathematical objects are formed by a collection of nodes and links connecting them [1]. Nodes represents physical objects at different size scales and the links joining them represent the physical relationships (interactions) existing between these objects. At the Planck-scale there are physical theories, like the loop quantum gravity [2], that predicts that space has a granular structure. Then, graphs in the form of *spin-networks* represent the space-time in a mathematically precise and physically compelling way. In such graphs nodes represent elementary grains of space which can be adjacent if they are separated by an elementary surface, which are represented by the links of the graph [2,3]. At a larger size scale, in nuclear and particle physics, graphs are used to represent particle interactions in the form of the *Feynman diagrams* [4]. Here the links represent *particle world lines* and nodes represent virtual interactions. There are two kinds of links called internal lines and external lines. The last are incident with only one node, and apart from them, a Feynman graph is analogous to a graph [4]. Thus, the vacuum-polarization graphs, which have no external lines are nothing but graphs. Increasing the size-scale we found the so-called *quantum graphs* [5,6], which are graphs in which lengths have been assigned to links and the graphs are equipped with self-adjoint differential or pseudo-differential operators. They include models for studying free electrons in organic molecules to studies of superconductivity in granular and artificial materials, waveguide networks, Anderson localization, mesoscopic quantum systems, and quantum chaos, among others [5,6].

Graph representation of chemical compounds have a large tradition in Chemistry and covers from small organic molecules to biomacromolecules and carbon nanotubes [7,8]. In these cases atoms are represented by the nodes of the graphs and (covalent) bonds are represented by links. More recently, the study of the so-called complex networks has received great attention in the scientific literature [9-13]. Complex networks are good examples of systems which pervades different scientific disciplines ranging from natural to technological and social sciences. The elements forming the system can be as diverse as routers in Internet, proteins in protein interaction networks or individuals in social networks [9-13]. Last but not least, a graph-theoretical picture of the universe at astronomical scale has been proposed to study galaxy distributions [14,15]. In this model galaxies are represented by the nodes of the *constellation graphs* and two nodes are linked if the corresponding galaxies are nearest neighbors. Thus, we find graph representing physical systems at scales ranged from $\sim 10^{-35}$ $m$ to $\sim 10^{22}$ $m$ and covering areas of study as diverse as quantum gravity, quantum electrodynamics, nano- and mesoscopic physic, econophysics and astrophysics.

Considering the widespread appearance of graphs in physical theories it is appealing to study new mathematical properties of these objects. These new findings can illuminate our understanding of the topological organization of our universe at different size scales. With this objective in mind we start here the investigation of a graph property which has not been previously recognized neither in mathematics nor in physics literature. It is based on graph spectral theory and deals with graphs having golden spectral ratio. The *golden ratio*, $\varphi = (1+\sqrt{5})/2$, which is also referred as the *golden mean* or *divine proportion*, pervades many different fields of science and arts [16-18]. The golden ratio plays a profound relation with nonlinear dynamics, chaos, fractals, knot theory, Cantorian spacetime, noncommutative geometry, quasi crystallography and geometry of four manifolds [19-22]. Recently, a golden rectangle has been used to derive the dilation of time intervals and the Lorentz contraction of lengths in special relativity [23]. A direct connection between Hilbert space and *E*-infinity theory has also been established via an irrational-transfinite golden ratio topological probability [24].

Here we prove the existence of graphs with golden spectral ratio. We elaborate some construction methods for this type of graphs and analyze some of their topological and dynamical properties, such as their expansibility and synchronizability. We also found some *almost* golden spectral graphs in nature, in particular in an ecological network representing trophic relations among species and in a literature citation network. The work is presented in a self-contained manner by introducing first the mathematical concepts used, then presenting the analytical results by means of Theorems and Propositions, and finally carrying out a computer search for golden spectral graphs in artificial and real-world networks.

**2. Preliminary definitions**

Let $G$ be a graph without loops or multiple links having $n$ nodes. Then the adjacency matrix of $G$, $\mathbf{A}(G) = \mathbf{A}$, is a square, symmetric matrix of order $n$, whose elements $A_{ij}$ are ones or zeroes if the corresponding nodes are adjacent or not, respectively. The sum of a row or column of this matrix is known as the degree of corresponding node and designated here by $d$ [1]. This matrix has $n$ (not necessarily distinct) real-valued eigenvalues, which are denoted here by $\lambda_1, \lambda_2, \ldots, \lambda_N$ [25]. The set of eigenvalues of $\mathbf{A}$ together with their multiplicities form the spectrum of $G$, which will be represented here as $Spec(G) = \{[\lambda_1]^{m_1}, [\lambda_2]^{m_2}, \cdots, [\lambda_n]^{m_n}\}$, where $\lambda_i$ is the $i$ th eigenvalue with $m_i$ multiplicity. Here the eigenvalues are usually assumed to be labelled in a non-increasing manner:

$$\lambda_1 \geq \lambda_2 \geq \cdots \geq \lambda_n. \tag{1}$$

Let $P_n$, $C_n$, $K_n$, $K_{a,b}$ $a+b=n$, $K_{n/c,n/c,\cdots,n/c}$ and $CP_n$ be the path graph, the complete graph, the complete bipartite graph, the complete multipartite graph with $n$ nodes and $c$ colours, and the cocktail-party graph, respectively [1]. The path $P_n$ is a tree with two nodes of degree 1, and the other $n-2$ nodes of degree 2. A tree is a graph without cycles. A cycle $C_n$ graph is a graph on $n$ nodes containing a single cycle through all nodes. $K_n$ is the graph in which every pair of nodes are connected. A bipartite graph is the one in which the set of nodes is decomposed into two disjoint sets such that no two nodes within the same set are adjacent. A complete bipartite graph $K_{a,b}$ is a bipartite graph such that every pair of nodes in the two sets are adjacent. If the nodes are partitioned into $k$ disjoint sets the graph is known as $k$-partite. Thus, $K_{n/c,n/c,\cdots,n/c}$ is the complete multipartite graph in which each partition contains $n/c$ nodes. The cocktail party graph $CP_n$, also called the hyperoctahedral graph, is the graph consisting of two rows of paired nodes in which all nodes but the paired ones are connected with a link.

Let $G$ and $G'$ be finite graphs with adjacency matrices $\mathbf{A}$ and $\mathbf{A}'$, respectively. Then we define the Kronecker (tensor) product of graph, $G \otimes G'$, as the graph whose adjacency matrix is $\mathbf{A} \otimes \mathbf{A}'$, where $\otimes$ is the Kronecker product of matrices defined by:

$$\mathbf{A} \otimes \mathbf{A}' = \begin{pmatrix} a_{11}\mathbf{A}' & a_{12}\mathbf{A}' & a_{13}\mathbf{A}' & \cdots \\ a_{21}\mathbf{A}' & a_{22}\mathbf{A}' & \cdots & \\ a_{31}\mathbf{A}' & \cdots & & \\ \vdots & & & \end{pmatrix}$$

The following matrices will be used for the tensor products, $\mathbf{J}_k$, $\mathbf{I}_k$ and $\mathbf{O}_k$, which represent the all-one, identity and all-zeroes matrices, respectively. Here we will use the notation used by van Dam [26-28] in which $G \tilde{\otimes} J_k$ represents the graph with adjacency matrix $(\mathbf{A}+\mathbf{I}_k) \otimes \mathbf{J}_k - \mathbf{I}_k$.

The line graph $L(G)$ of a graph $G$ is obtained by associating a node with each link of $G$ and connecting two nodes with a link in $L(G)$ if and only if the corresponding links of $G$ have a common node.

## 3. Spectral ratios

We start by defining two new measures for graphs based on the spectra of adjacency matrices. Let $\lambda_1 - \lambda_n$ be the *length* of the spectrum of $G$ and $\lambda_1 - \lambda_2$ its *spectral gap*. Then, we define the *width* of the "bulk" part of the spectrum as $\lambda_2 - \lambda_n$. Using these measures we define the spectral ratios of the graph $G$ as the proportion between width of the "bulk" part of the spectrum and the spectral gap, as well as the ratio between spectral length and the width of the "bulk" part of the spectrum:

$$w_1(G) = \frac{\lambda_2 - \lambda_n}{\lambda_1 - \lambda_2} \qquad (\lambda_1 - \lambda_2) \neq 0 \qquad (2)$$

$$w_1(G) = \frac{\lambda_1 - \lambda_n}{\lambda_2 - \lambda_n} \qquad (\lambda_2 - \lambda_n) \neq 0 \qquad (3)$$

Let consider the particular case in which $w_1(G) = w_2(G)$. That is,

$$\frac{\lambda_2 - \lambda_n}{\lambda_1 - \lambda_2} = \frac{\lambda_1 - \lambda_n}{\lambda_2 - \lambda_n} = \varphi \qquad (4)$$

Then, we have that $\lambda_2 - \lambda_n = \varphi(\lambda_1 - \lambda_2)$, which can be substituted on the left part of (4) giving

$$\frac{(\lambda_1 - \lambda_2) + \varphi(\lambda_1 - \lambda_2)}{\varphi(\lambda_1 - \lambda_2)} = \frac{\varphi(\lambda_1 - \lambda_2)}{\lambda_1 - \lambda_2} \qquad (5)$$

which is reduced to $\varphi^2 - \varphi - 1 = 0$ where $\varphi = \frac{1 + \sqrt{5}}{2}$ is the *golden ratio, golden mean* or *divine proportion*.

Let $G$ be a graph with spectral ratios defined by (4). Then, we call this graph a *graph with golden spectral ratio* or *golden spectral graph* (GSG) for short. GSGs will have several interesting mathematical properties related to their spectral ratios. For instance, their spectral ratios [16-18]:

i) can be expressed as a self-similar continued fraction

$$\frac{\lambda_2 - \lambda_n}{\lambda_1 - \lambda_2} = \frac{\lambda_1 - \lambda_n}{\lambda_2 - \lambda_n} = 1 + \cfrac{1}{1 + \cfrac{1}{1 + \cfrac{1}{1 + \cdots}}} \tag{6}$$

ii) can be obtained as the sum of all its reciprocal powers,

$$\frac{\lambda_2 - \lambda_n}{\lambda_1 - \lambda_2} = \frac{\lambda_1 - \lambda_2}{\lambda_2 - \lambda_n} + \left(\frac{\lambda_1 - \lambda_2}{\lambda_2 - \lambda_n}\right)^2 + \left(\frac{\lambda_1 - \lambda_2}{\lambda_2 - \lambda_n}\right)^3 + \cdots \tag{7}$$

iii) are the most irrational number, which is exactly the value at which the sequence of ratios of consecutive Fibonacci numbers converges

$$\frac{\lambda_2 - \lambda_n}{\lambda_1 - \lambda_2} = \frac{\lambda_1 - \lambda_n}{\lambda_2 - \lambda_n} = \lim_{n \to \infty} \frac{F_n}{F_{n-1}} \tag{8}$$

## 4. The smallest graph with golden spectral ratio

In this section we are interested in finding the smallest connected graph with golden spectral ratio. We first prove the following result for cycle graphs.

**Theorem 1**. Let $C_n$ be the cycle graph of $n$ nodes. Then, $C_n$ is a GSG if, and only if, $n = 5$.

**Proof**. The spectra of $C_n$ is given by [25]:

$$SpecC_n = \left([2]^1, [2\cos 2\pi/n]^2, \cdots, [2\cos(n-1)\pi/n]^2\right) \quad (n \text{ odd}) \tag{9}$$

$$SpecC_n = \left([2]^1, [2\cos 2\pi/n]^2, \cdots, [2\cos(n-2)\pi/n]^2, [-2]^1\right) \quad (n \text{ even}) \tag{10}$$

Let represent $C_n$ as a regular $n$-gon with edge length $a_n = 2R_n \sin(\pi/n)$, where $R_n$ is the circumradius. Then,

$$\lambda_2 = 2 - (a_n/R_n)^2 \qquad (11)$$

$$\lambda_n = -\frac{\sqrt{4R_n^2 - a_n^2}}{R_n} \quad (n \text{ odd}) \qquad (12)$$

From where we can obtain the expressions for the spectral ratios

$$w_1(C_n) = \frac{2R_n^2 - a_n^2 + R_n\sqrt{4R_n^2 - a_n^2}}{a_n^2} \quad (n \text{ odd}) \qquad (13)$$

$$w_1(C_n) = \frac{2R_n^2 - a_n^2}{a_n^2} \quad (n \text{ even}) \qquad (14)$$

Then, for $a_n = 1$, and using the relation between the circumradius and $a_n$ we obtain for a cyclic graph with golden spectral ratio:

$$n = \frac{\pi}{\sin^{-1}\left(\frac{\sqrt{4\varphi^2 - 1}}{2\varphi^2}\right)} \quad (n \text{ odd}) \qquad (15)$$

$$n = \frac{\pi}{\sin^{-1}(1/\varphi)} \quad (n \text{ even}) \qquad (16)$$

Then, we obtain $n = 5$ and $n = 4.72$ (not an integer number) from expressions (15) and (16), respectively, which proves that $C_5$ is the only cyclic graph with golden spectral ratio.

This can be observed graphically by plotting both spectral ratio measures versus the number of nodes in the cycle graphs. In Fig. 1a we can see that both spectral ratio measures intersect to each other at exactly the value of $n = 5$, which corresponds to the golden ratio.

**Insert Fig. 1 about here.**

The following result shows that the pentagon is the simplest connected graph displaying a golden spectral ratio.

**Proposition 1**. $C_5$ is the smallest connected graph with golden spectral ratio.

**Proof**. We simply calculate $w_1(G)$ for all connected graphs with $n$ vertices, $2 \leq n \leq 5$, for which no one graph is a GSG.

## 5. Non-existence results

It is interesting to investigate where not to search for graphs with GSR. Here we find two results which exclude large families of graphs in the search for GSGs. The first result includes five different families of graphs, which are commonly found in several applications of graphs in physical contexts.

**Theorem 2**. For each $n$ there is no GSGs among $P_n$, $K_n$, $K_{a,b}$ $a+b=n$, $K_{n/c,n/c,\cdots,n/c}$, $CP_n$.

**Proof**. It is not difficult to see that $w_1(K_n) = 0$, $w_2(K_n)$ is undefined, $w_1(K_{a,b}) = 1$ and $w_2(K_{a,b}) = 2$. The spectrum of $K_{n/c,n/c,\cdots,n/c}$ is $Spect(K_{n/c,n/c,\cdots,n/c}) = \{[n-n/c]^1, [0]^{n-c}, [-n/c]^{c-1}\}$ [29]. Thus, $w_1(K_{n/c,n/c,\cdots,n/c}) = 1/(c-1)$ and $w_2(K_{n/c,n/c,\cdots,n/c}) = c$. For the cocktail-party graph we have $Spect(CP_n) = \{[2n-2]^1, [0]^n, [-2]^{n-1}\}$ [25,29] and we obtain $w_1(CP_n) = 2/(2n-2)$ and $w_2(CP_n) = n$. Thus, it is enough to see that $w_1(G)$ is bounded between 0 and 1 for these two types of graphs $0 \leq w(K_{n/v,n/v,\cdots,n/v}) \leq 1$ and $0 \leq w_1(CP_n) \leq 1$. The path $P_n$ has eigenvalues $\lambda_j = 2\cos\left(\dfrac{j\pi}{n+1}\right)$ for $j = 1,2,\cdots,n$ [29]. Then it is easy to see that $w_1(P_n)$ and $w_2(P_n)$ never intersect to each other (see Fig. 1b). Note that $w_2(P_2)$ is undefined and that there is no sense in calculating the spectral ratios for $P_1$ as it has only one eigenvalue.

The second non-existence results is related to the graphs of symmetric balanced incomplete block designs [30,31]. Combinatorial designs have been largely applied in the statistical design of experiments and in the theory of error-correcting codes [30,31]. More recently they have found a large number of applications in computer science and the theory of cryptography and networking [32]. Let $G$ be a symmetric balanced incomplete block design consisting of $v$ elements and $B$ blocks, such that each element is contained in $d$ blocks, each block contains $d$ elements and each pair of elements is simultaneously contained in $\lambda*$ blocks. In this case we have the following non-existence result.

**Theorem 3**. For each $v, d, \lambda*$ there is no incidence graph of symmetric $2-(v, d, \lambda*)$ designs which is GSGs.

**Proof**. The spectrum of the incidence graph of symmetric $2-(v, d, \lambda*)$ designs is [29]

$$\left\{ [d]^1, [\sqrt{d-\lambda*}]^{v-1}, [-\sqrt{d-\lambda*}]^{v-1}, [-d]^1 \right\} \tag{17}$$

Thus the spectral ratio for these graph is

$$w_1(G) = \frac{d + \sqrt{d-\lambda*}}{d - \sqrt{d-\lambda*}} \tag{18}$$

For the case of $w_1(G) = \varphi$ we obtain the following relation between the number of blocks and the number of elements per block:

$$\lambda* = -d\left[ d\frac{(\varphi-1)^2}{\varphi^4} - 1 \right] \tag{19}$$

Due to their definitions both the number of blocks and the number of elements per block should be integer positive numbers. It is easy to see that $\lambda* < 0$ for $d \geq 18$. On the other hand, there is no integer value of $\lambda*$ for $0 < d < 18$, which proves the previous result.

**Remark**. There is one block design for which the spectral ratio is approximated to the golden ratio. It corresponds to the design with $d = 6$, $\lambda^* = 4$ ($w_1(G) \approx 1.617$).

## 6. Construction methods

In this section we propose three methods for building graphs which have golden spectral ratios. The first two cases generate infinite families of GSGs and the last case generate GSGs at the infinite size limit. The first of these families is built by using the following result.

**Theorem 4**. The graphs $C_5 \otimes J_k$ are GSGs.

**Proof**. If $G$ has $n$ nodes and spectrum $\{[r]^1, [p]^f, [0]^m, [s]^g\}$, where $m$ could be zero, then $G \otimes J_k$ has $nk$ nodes and spectrum $\{[rk]^1, [pk]^f, [0]^{m+nr-n}, [sk]^g\}$. Thus, $C_5 \otimes J_k$ has spectrum [26-28]

$$Spec(G) = \left\{[2k]^1, \left[-\frac{k}{2} + \frac{k}{2}\sqrt{5}\right]^2, [0]^{5(k-1)}, \left[-\frac{k}{2} - \frac{k}{2}\sqrt{5}\right]^2\right\} \tag{20}$$

from which we obtain $w_1(G) = \dfrac{2\sqrt{5}}{5 - \sqrt{5}} = \varphi$.

Now let consider the $k$-covers of $C_3 \otimes J_k$ and $C_5 \otimes J_k$. Let $\mathbf{C}$ be the $k \times k$ circulant matrix whose elements are $C_{ij} = 1$ if $j = i + 1 \pmod{k}$, and $C_{ij} = 0$ otherwise. Then let $\mathbf{P}$ be the $k^2 \times k^2$ matrix defined as follow [26]

$$\mathbf{P} = \begin{pmatrix} \mathbf{I} & \mathbf{I} & \cdots & \mathbf{I} \\ \mathbf{C} & \mathbf{C} & \cdots & \mathbf{C} \\ \vdots & \vdots & & \vdots \\ \mathbf{C}^{k-1} & \mathbf{C}^{k-1} & \cdots & \mathbf{C}^{k-1} \end{pmatrix}.$$

In addition, let define $\mathbf{D} = (\mathbf{J}_k - \mathbf{I}_k) \otimes \mathbf{I}_k$. Then, the $k$-covers of $C_3 \otimes J_k$ and $C_5 \otimes J_k$ will have the following adjacency matrices [26]

$$\mathbf{A}_3 = \begin{pmatrix} \mathbf{D} & \mathbf{P} & \mathbf{P}^T \\ \mathbf{P}^T & \mathbf{D} & \mathbf{P} \\ \mathbf{P} & \mathbf{P}^T & \mathbf{D} \end{pmatrix}, \text{ and } \mathbf{A}_5 = \begin{pmatrix} \mathbf{D} & \mathbf{P} & 0 & 0 & \mathbf{P}^T \\ \mathbf{P}^T & \mathbf{D} & \mathbf{P} & 0 & 0 \\ 0 & \mathbf{P}^T & \mathbf{D} & \mathbf{P} & 0 \\ 0 & 0 & \mathbf{P}^T & \mathbf{D} & \mathbf{P} \\ \mathbf{P} & 0 & 0 & \mathbf{P}^T & \mathbf{D} \end{pmatrix}.$$

**Theorem 5.** The $k$-covers of $C_3 \otimes J_k$ and $C_5 \otimes J_k$ are GSG.

**Proof.** The spectra of the $k$-covers of $C_3 \otimes J_k$ and $C_5 \otimes J_k$ are, respectively [26-28],

$$Spect(G) = \left\{ [3k-1]^1, [-1]^{3k^2-6k+5}, \left[-1+k\frac{1\pm\sqrt{5}}{2}\right]^{3k-3} \right\} \tag{21}$$

$$Spect(G) = \left\{ [3k-1]^1, [-1]^{5k^2-10k+5}, \left[-1+k\frac{1\pm\sqrt{5}}{2}\right]^{5k-3} \right\} \tag{22}$$

From which,

$$w_1(G) = \frac{2\sqrt{5}}{5-\sqrt{5}} = \varphi$$

**Corollary.** The icosahedral graph is a GSG.

The icosahedral graph is the 2-cover of $C_3 \otimes J_2$, from which we deduce that it is GSG according to Theorem 5. We have checked by computer that the icosahedron is the only GGST among the platonic graphs, i.e., tetrahedral, cubical, octahedral, icosahedral and dodecahedral graphs.

The final result in this section is concerned to the line graph of complete bipartite graphs.

**Theorem 6.** The line graph of the complete bipartite graph, $L(K_{a,b})$ for which $a = F_{k+1}$, $b = F_k$, where $F_k$ is the $k$th Fibonacci number is a GSG for $k \to \infty$.

**Proof.** The spectrum of $L(K_{a,b})$ for $a > b \geq -2$ is [29]:

$$Spect[L(K_{a,b})] = \{[a+b-2]^1, [a-b]^{b-1}, [b-2]^{a-1}, [-2]^{ab-a-b+1}\}. \tag{23}$$

Then,

$$w_1[L(K_{a,b})] = \frac{a}{b} = \frac{F_{k+1}}{F_k} \quad \text{and} \quad w_2[L(K_{a,b})] = \frac{a+b}{a} = \frac{F_k + F_{k+1}}{F_{k+1}} \tag{24}$$

It is known [33] that $\lim_{k \to \infty} \frac{F_{k+1}}{F_k} = \varphi$ and $\lim_{k \to \infty} \frac{F_k + F_{k+1}}{F_{k+1}} = \varphi$ which proves the result.

**Remark**. The theorem 6 is also true for $a = L_{k+1}$ and $b = L_k$ where $L_k$ is the $k$ th Lucas number [33].

The graphs generated by the three construction methods developed here are $d$-regular graphs. The graphs generated by Theorem 4 have $n = 5k$ nodes and degree equal to $2k$, i.e., they are $2k$-regular graphs. The graphs generated by Theorem 5 have $n = 3k^2$ and $n = 5k^2$ nodes, respectively and are $(3k-1)$-regular graphs. In the last case (Theorem 6) the graphs generated will have $n = ab$ nodes and they will be $(a+b-2)$-regular graphs.

## 7. Expansibility of GSGs

Expansibility is an interesting property that deserves to be investigated for GSGs. Good expansion networks (GENs) show excellent communication properties due to the absence of bottlenecks [34]. A bottleneck is a small set of nodes/links whose elimination leads to fragmentation of the network into at least two large connected components. Formally, a graph is considered to have GE if every subset $S$ of nodes ($S \leq 50\%$ of the total number of nodes) has a neighborhood that is larger than some "expansion factor" $\phi$ multiplied by the number of nodes in $S$. A neighborhood of $S$ is the set of nodes which are linked to the nodes in $S$. Formally, for each vertex $v \in V$ (where $V$ is the set of nodes in the network), the neighborhood of $v$, denoted as $\Gamma(v)$ is defined as: $\Gamma(v)=\{u \in V | (u,v) \in E\}$ (where $E$ is the set of links in the network). Then, the neighborhood of a subset $S \subseteq V$ is defined as the union of the neighborhoods of the nodes in $S$: $\Gamma(S) = \bigcup_{v \in S} \Gamma(v)$ and the network has GE if $\Gamma(v) \geq \phi|S|\ \forall S \subseteq V$.

Let $\Delta = \lambda_1 - \lambda_2$ be the *spectral gap* of a graph. Then it is known that for a $d$-regular graph the expansion parameter is related to the spectral gap by [34]:

$$\frac{\Delta}{2} \leq \phi \leq \sqrt{2d\Delta} \tag{25}$$

Thus, the larger the spectral gap the larger the expansion of the graph. Among the graphs with large spectral gap there is a family of graph named Ramanujan graphs [35,36]. These graphs have spectral gaps almost as large as possible. Formally, a Ramanujan graph is a $d$-regular graph for which [35,36]

$$\lambda(G) \leq 2\sqrt{d-1} \tag{26}$$

where $\lambda(G)$ is the maximum of the non-trivial eigenvalues of the graph

$$\lambda(G) = \max_{|\lambda_i| < d} |\lambda_i| \tag{27}$$

The graphs $C_5 \otimes J_k$ and the $k$-covers of $C_3 \otimes J_k$ and $C_5 \otimes J_k$ have spectral gaps given by the following expressions, respectively

$$\Delta_1 = \left(\frac{1}{\varphi^2 + 1}\right)n,\ \Delta_2 = \left(\frac{\varphi^2 + 1}{\sqrt{3}}\right)\sqrt{n},\ \text{and}\ \Delta_3 = \left(\frac{\varphi^2 + 1}{\sqrt{5}}\right)\sqrt{n} \tag{28}$$

Consequently, we have the following results for these GSGs.

**Proposition 2**. The graphs $C_5 \otimes J_k$ are Ramanujan for $k \leq 20$.

**Proof**. The largest non-trivial eigenvalue for $C_5 \otimes J_k$ is $\lambda_2 = \dfrac{k}{\varphi}$. Thus we have that these graphs are Ramanujan if, and only if, $\dfrac{k^2}{\varphi^2} - 8k + 2 \leq 0$, which is true for $k \leq 20$ ($n \leq 100$).

**Proposition 3**. The *k*-covers of $C_3 \otimes J_k$ and $C_5 \otimes J_k$ are Ramanujan for $k \leq 9$.

**Proof**. The largest non-trivial eigenvalue for $C_3 \otimes J_k$ and $C_5 \otimes J_k$ is $\lambda_2 = \varphi k - 1$. Then, these graphs are Ramanujan if, and only if, $\varphi^2 k^2 - 2k(6 + \varphi) + 5 \leq 0$, which is true for $k \leq 9$, $n \leq 265$ for $C_3 \otimes J_k$ and $n \leq 405$ for $C_3 \otimes J_k$.

These results indicate that we have a total of 20 Ramanujan graphs among the graphs $C_5 \otimes J_k$. These graphs are relatively small having between 5 and 100 nodes. On the other hand, there are 18 Ramanujan graphs among $C_3 \otimes J_k$ and $C_5 \otimes J_k$ which have up to 405 nodes. In Fig. 2 we illustrate some of the Ramanujan graphs constructed by using Propositions 2 and 3.

**Insert Fig. 2 about here.**

In general we can show the following general result for a $d$-regular graph.

**Proposition 4**. A $k$-regular graph is GSG if and only if

$$\lambda_2 > \frac{d}{\varphi^3} \tag{29}$$

**Proof**. Let $d = \lambda_1 > \lambda_2 \geq \cdots \geq \lambda_n \geq -d$ be the eigenvalues of the $d$-regular graph and let $\lambda_n = -d + \varepsilon$ for $0 \leq \varepsilon < 2d$. Then, the graph is GSG if and only if

$$\lambda_2 = \frac{d(\varphi - 1) + \varepsilon}{\varphi + 1} \tag{30}$$

Thus $\lambda_2 > \dfrac{d(\varphi-1)}{\varphi+1} = \dfrac{d}{\varphi^3}$.

This proposition have three important consequences which are directly related to the good expansion properties of GSGs because of the connection between Ramanujan graphs and expanders [34-36]. They are,

i) No $d$-regular GSG is Ramanujan for $d > 70$,

ii) Every bipartite $d$-regular GSG is Ramanujan for $d \leq 70$,

iii) No graph generated by Theorem 6 is Ramanujan.

These consequences are easily derived from the definition of Ramanujan graphs in combination with proposition 4. A $d$-regular GSG is Ramanujan if

$$\dfrac{d}{\varphi^3} \leq 2\sqrt{d-1} \qquad \text{or} \qquad d(d-4\varphi^6)+4\varphi^6 \leq 0 \qquad (31)$$

which is satisfied for $d \leq 70$. Because $\varepsilon = 0$ is the lower bound for this constant, which corresponds to the bipartite $d$-regular graph it is easy to see that every bipartite $d$-regular GSG is Ramanujan for $d \leq 70$. The graphs generated by Theorem 6 have degree much larger than 70 due to the fact that they are built for the infinite limit. Thus it is obvious that they are not Ramanujan. This can be seen graphically in Fig. 3 where the area demarked between the curves $\lambda_2 = 2\sqrt{d-1}$ and $\lambda_2 = \dfrac{d}{\varphi^3}$ corresponds to the region where Ramanujan $d$-regular GSGs can be found. All Ramanujan regular GSGs previously found in this work are located in this region. All these graphs display excellent communication properties and well as large robustness due to the lack of structural bottlenecks, which make them good candidates for infrastructure and communication networks [34].

**Insert Fig. 3 about here.**

## 8. Synchronizability of GSGs

Now we are going to analyze the synchronizability of individual dynamical processes occurring at the vertices of a graph. We will consider the criterion established by Barahona and Pecora based on spectral techniques to determine the stability of synchronized states on networks [37]. These results have been used by Donetti et al. [38,39] for finding entangled networks, which are displayed to have super-homogeneity and optimal topologies. In this approach a dynamical process $\dot{x}_i = F(x_i) - \sigma \sum_k L_{ij} H(x_j)$ is considered. In this case the dynamical variables are $x_{i=1,2,...,n}$. $F$ and $H$ are the evolution and the coupling functions, respectively, $\sigma$ is a constant and $L_{ij}$ are the entries of the discrete Laplacian matrix $\mathbf{L}(G)$, which is defined as follows [25]:

$$\mathbf{L} = \mathbf{L}(G) = \mathbf{D} - \mathbf{A} \tag{32}$$

where $\mathbf{D}$ is the degree matrix whose diagonal entries are the degrees of the corresponding nodes and the rest of entries are zeroes, and $\mathbf{A}$ is the adjacency matrix.

According to the approach followed by Barahona and Pecora [37] it is concluded that a graph exhibits *better synchronizability* if the ratio $Q = \dfrac{\mu_1}{\mu_{n-1}}$ is as small as possible, where $\mu_j$ is an eigenvalue of the Laplacian matrix, which have been ordered in a non-increasing manner: $\mu_1 > \mu_2 \geq \ldots \geq \mu_{n-1} \geq \mu_n = 0$.

It is easy to see that for regular graphs the following relationship exists between the eigenvalues of the Laplacian and the eigenvalues of the adjacency matrix of a graph (ordered as in (1)):

$$\mu_j = \lambda_1 - \lambda_{n-j+1} \tag{33}$$

Thus, it is straightforward to realize that the ratio $Q$ can be expressed in terms of the adjacency eigenvalues of a $d$-regular graph as follows

$$Q = \frac{\lambda_1 - \lambda_n}{\lambda_1 - \lambda_2} = \frac{d - \lambda_n}{d - \lambda_2} \tag{34}$$

Then, for a $d$-regular graph $Q$ is simply the product of both spectral ratios defined in this work: $Q = w_1(G)w_2(G)$, which immediately implies that for a $d$-regular GSG: $Q = \varphi^2 = \varphi + 1$.

Using an optimization process based on minimizing the ratio $Q$, Donetti et al. have found the so-called "entangled" networks, which are graphs having low values of $Q$ and thus displaying good synchronizability and good expansion properties [38,39]. Among the graphs found by this process there are some well-known graphs, such as the 3-cages with 10, 14 and 24 nodes, which are known as *Petersen*, *Heawood* and *McGee* graphs. Other entangled graphs which are not $d$-cages were also reported by these authors and they are shown in Fig. 4 together with the 3-cages.

**Insert Fig. 4 about here.**

The $Q$ ratios for these optimal graphs found by Donetti et al. [38,39] are 2.500, 2.784, 5.562, 3.613 and 4.796 for the Petersen, Heawood, McGee graphs and optimal entangled networks with $d = 3$ with 12 and 16 nodes, respectively. As can be seen only the Petersen graph has $Q < \varphi^2$, $Q < Q_{GSG}$. We can call these graphs having synchronizability better than a GSG a *platinum* graph. We speculate that the number of platinum graphs is very small and consequently GSGs are among the graphs with lowest $Q$ ratios and better synchronizability that exist.

## 9. Computer results

Now we are going to explore some series of graphs for which we have not obtained analytic results in order to search for GSGs. These sets of graphs includes all trees with $n$ nodes, $2 \leq n \leq 10$, all connected cubic graphs with $n$ nodes for $n = 4,6,8,10,12$ and a set of miscellaneous graphs [40].

The computer analysis of 200 trees does not provide any GSG. There is one graph which displays values of the spectral ratios which are close to the golden ratio (see Fig. 5). This graph corresponds to a class of graphs known as comets [41]. A comet, $Cm(q,r)$, having $n = q + r + 1$ nodes is a graph formed by a star $K_{1,q}$ and a path $P_r$ directly bonded to one of the nodes of the star which is different from the central one:

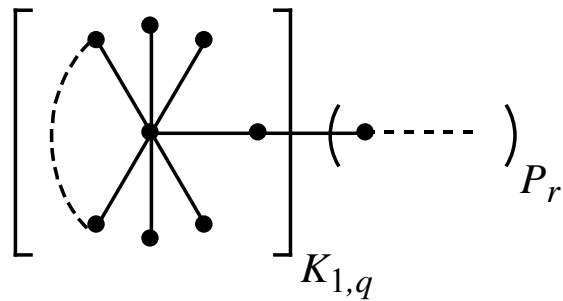

**Insert Fig. 5 about here.**

The graph we found to have spectral ratio close to the golden section corresponds to the comet $Cm(8,1)$, which has $w_1(G) = 1.963$ and $w_2(G) = 1.510$. Then, we explore all comets $Cm(q,1)$, $3 \leq q \leq 20$ ($5 \leq n \leq 22$). The results are illustrated in Fig. 6 where it can be seen that both spectral ratios are very close to each other for the comet $Cm(17,1)$ which has $w_1(G) = 1.6125$ and $w_2(G) = 1.6202$. Thus, this comet can be considered as an *almost* GSG.

**Insert Fig. 6 about here.**

The second family of graphs whicht is analysed here is that of cubic graphs. In Fig. 5 we illustrate the trends of both spectral ratios as a function of the spectral gap. As can be seen they follows similar trends than the ones observed from trees. Here there are three graphs which have spectral ratios close to the golden section. The one which is closest to a GSG is a circulant graph with 8 vertices. A circulant graph with $n$ nodes [25], $Ci_n(j)$, is a graph in which the $i$th node is adjacent to the $(i+j)$th and $(i-j)$th nodes for each $j$ in a list $l$. The graph we have identified to be almost GSG is $Ci_8(4)$, which has $w_1(G)=1.705$ and $w_2(G)=1.5865$. Then we have explored all circulant graph of the type $Ci_n(n/2)$ for $n \leq 12$ and $n$ even. We have seen that both spectral ratios diverge as the number of nodes increases and the graph which is closest to a GSG is $Ci_8(4)$. Another graph in this family was also found among the ones having spectral ratios close to the golden section. It is $Ci_6(3)$ having $w_1(G)=1.500$ and $w_2(G)=1.667$. With identical spectral ratios we also found another cubic graph, which corresponds to the Petersen graph [42]. Consequently, we have explored the generalized Petersen graphs with $n=6,8,10,12,14$ nodes and we found that among them the "classical" Petersen graph has the spectral ratios closest to the golden section.

Finally, we explored 27 miscellaneous graphs for which we plot both spectral ratios as a function of the spectral gap in Fig. 5. Among these graphs we found 4 which are *almost*-GSGs and which are illustrated in Fig. 7. The spectral ratios for these graphs are, according to Fig. 5: $w_1(a)=1.5757$, $w_2(a)=1.6346$, $w_1(b)=1.5583$, $w_2(b)=1.6417$, $w_1(c)=1.6916$, $w_2(c)=1.5912$, $w_1(d)=1.6986$, $w_2(d)=1.5887$. These graphs are examples of not regular graphs which are *almost* GSGs. We hope that further investigation of larger pools of graphs will identify a larger number of GSGs and almost-GSGs.

**Insert Fig. 7 about here.**

## 10. Real-world complex networks

We study here 47 real-world complex networks accounting for ecological, biological, informational, technological and social systems. The ecological networks studied correspond to the following food webs [43]: Benguela, Bridge Brook, Canton

Creek, Chesapeake Bay, Coachella Valley, El Verde rainforest, Grassland, Little Rock Lake, Reef Small, Scotch Broom, Shelf, Skipwith Pond, St. Marks Seagrass, St. Martin Island, Stony, Ythan Estuary (1) with and without parasites (2). The biological networks correspond to the protein–protein interaction networks (PINs), for *Saccharomyces cerevisiae* (yeast) and for the bacterium *Helicobacter pylori*; three transcription interaction networks concerning *E. coli*, yeast and sea urchins and the neural network in *C. elegans*. The informational networks include two semantic networks, one based on Roget's Thesaurus of English (Roget) and the other on the Online Dictionary of Library and Information Science (ODLIS) and four citation networks: one consisting of papers published in the *Proceedings of Graph Drawing* in the period 1994–2000 (GD), papers published in the field of "Network Centrality" (Centrality), papers published or citing articles from *Scientometrics* for the period 1978–2000 (SciMet) and papers containing the phrase "Small World". The technological systems represented by networks correspond to three electronic sequential logic circuits parsed from the ISCAS89 benchmark set, where nodes represent logic gates and flip-flops, the airport transportation network in the US in 1997, the Internet at the autonomous systems (AS) level as from April 1998 and five software networks: Abi, Digital, MySQL, VTK and XMMS. Finally, the social networks studied here include a network of the corporate elite in US, a scientific collaboration network in the field of computational geometry (Geom), inmates in prison, injectable drug users (IDUs), Zachary karate club, college students on a course about leadership, and a collaboration between Jazz musicians.

We calculated the spectral ratios for these networks in search for GSG or almost-GSGs. The results are plotted in Fig. 5 as a function of the spectral gap. As can be seen the same trend observed for the different classes of graphs previously analysed is also observed for these real-world complex networks. That is, the largest the spectral gap the lowest the first spectral ratio, $w_1(G)$, while the values of $w_2(G)$ are mainly concentrated between 1 and 2.

There is not one real-world network with golden spectral ratio but two of them can be considered as almost-GSG. They correspond to the food web of St. Marks National Refuge, in Florida, USA and the other to the citation network for papers published in the field of "Network Centrality" (see Fig. 8). The first network has 48 nodes and average degree $\langle k \rangle = 9.08$. It has spectral ratios of $w_1(G) = 1.630$ and $w_2(G) = 1.613$. The second almost-GSR network has 118 nodes, $\langle k \rangle = 10.4$, $w_1(G) = 1.655$ and $w_2(G) = 1.604$. These networks have been previously identified to have large spectral gaps and displaying good expansion properties. The St. Marks food web also has an uniform degree distribution implying large robustness due to the absence of structural bottlenecks and hubs connecting large number of nodes. The Centrality network display an exponential degree distribution and can be vulnerable by attacking the most connected nodes in the network. Both networks are also expected to display very low $Q$ ratio and consequently they should display good synchronizability.

**Insert Fig. 8 about here.**

**11. Conclusions**

We have investigated here the existence of graph (networks) displaying golden spectral ratios. The spectral ratios have been introduced here in a very intuitive way by measuring the proportions between the width of the "bulk" part of the spectrum and the spectral gap, as well as the ratio between spectral length and the width of the "bulk" part of the spectrum. We have proved that the pentagon is the smallest golden spectral graph (GSG). It is the only cycle graph with this property and it should be mention that a regular pentagon also display a golden ratio between the length of its diagonal and the length of its edges. Thus, it is both topologically and geometrically golden. We have proposed three construction methods that permit to build families of GSGs. We have shown that some of the GSGs built using these results display very interesting topological and dynamical properties. For instance, several of these graphs are Ramanujan, which make them to display very good expansion properties. We have also investigated the synchronizability of GSGs and we have found that these graphs are among the one showing the best possible synchronizability. In consequence, GSGs are very good candidates to design robust networks for different technological, social and infrastructure systems.

As normally happens when a new property is discovered there are more open questions than answers. The investigation of golden spectral graphs is in this stage and we consider that a great effort need to be done to understand all mathematical and physical consequences of a graph displaying golden spectral properties. We hope, however, that the current work contribute to motive this research in both mathematics and physics.

**Acknowledgement**

The author thanks the program "Ramón y Cajal", Spain for partial financial support.

# Figure Captions

Fig. 1. Plot of spectral gaps, $w_1(G)$, continuous line and $w_2(G)$ discontinuous line, versus the number of nodes in linear-logarithmic scale for (a) cycle graphs, $C_n$ and (b) path graphs, $P_n$. The dot line represents the value of $\varphi = (1+\sqrt{5})/2$.

Fig. 2. Some golden spectral graphs which are also Ramanujan graphs.

Fig. 3. Plot of the second largest eigenvalue of the adjacency matrix, $\lambda_2$, for $d$-regular graphs as a function of node degree, $d$. Ramanujan golden spectral graphs are those between the continuous, $2(d-1)^{0.5}$, and discontinuous, $d/\varphi^3$, lines.

Fig. 4. Some of the entangled graphs found by Donetti et al. (see text for explanation).

Fig. 5. Plot of spectral ratios, $w_1(G)$, circles and $w_2(G)$ squares, as a function of the spectral gap $\lambda_1 - \lambda_2$ for trees with $2 \leq n \leq 10$, cubic graphs with $n = 4,6,8,10,12$, miscellaneous graphs and real-world networks. The discontinuous line represents the value of $\varphi = (1+\sqrt{5})/2$.

Fig. 6. Plot of spectral gaps, $w_1(G)$, continuous line and $w_2(G)$ discontinuous line, versus the number of nodes in linear-logarithmic scale for comet graphs The dot line represents the value of $\varphi = (1+\sqrt{5})/2$. Both spectra ratio curves intersect close to the value of $n = 19$.

Fig. 7. Some miscellaneous graphs which are found to be almost golden spectral graphs (see text for the values of the spectral ratios).

Fig. 8. Two real-world networks with spectral ratios very close to the golden ratio.

Fig. 1

(a)

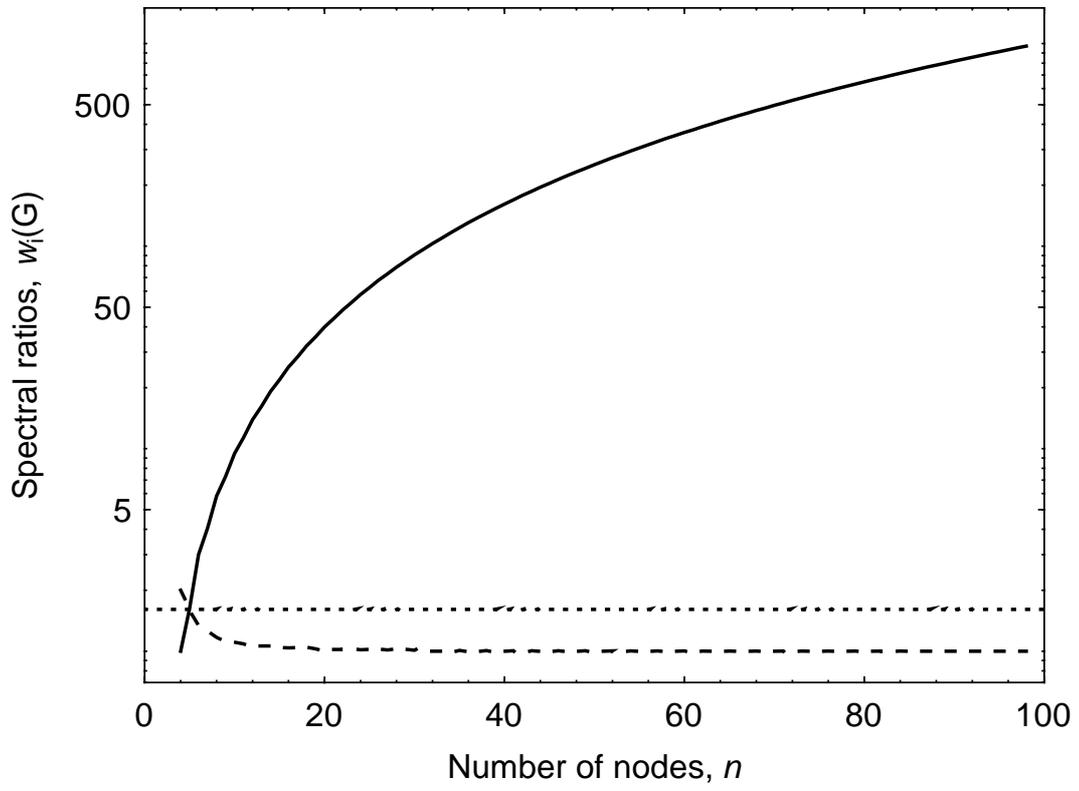

(b)

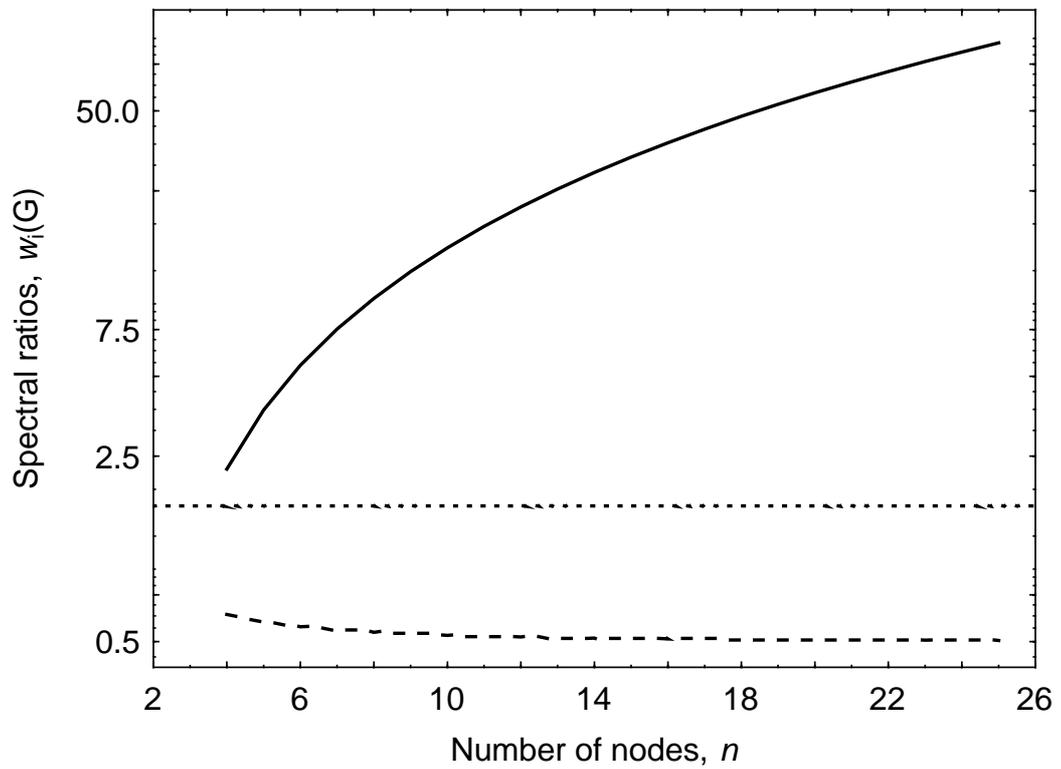

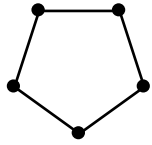
$C_5$

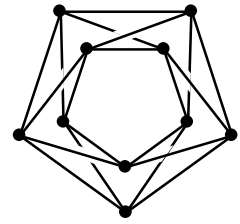
$C_5 \otimes J_2$

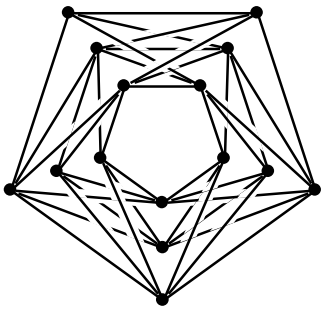
$C_5 \otimes J_3$

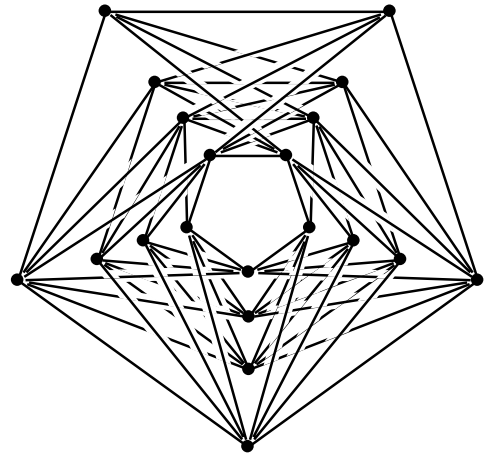
$C_5 \otimes J_4$

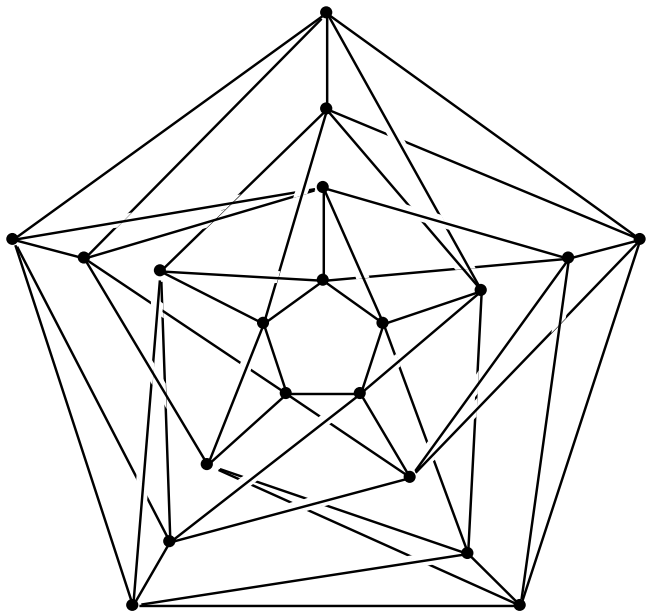
2-cover of $C_5 \circledast J_2$

Fig. 3

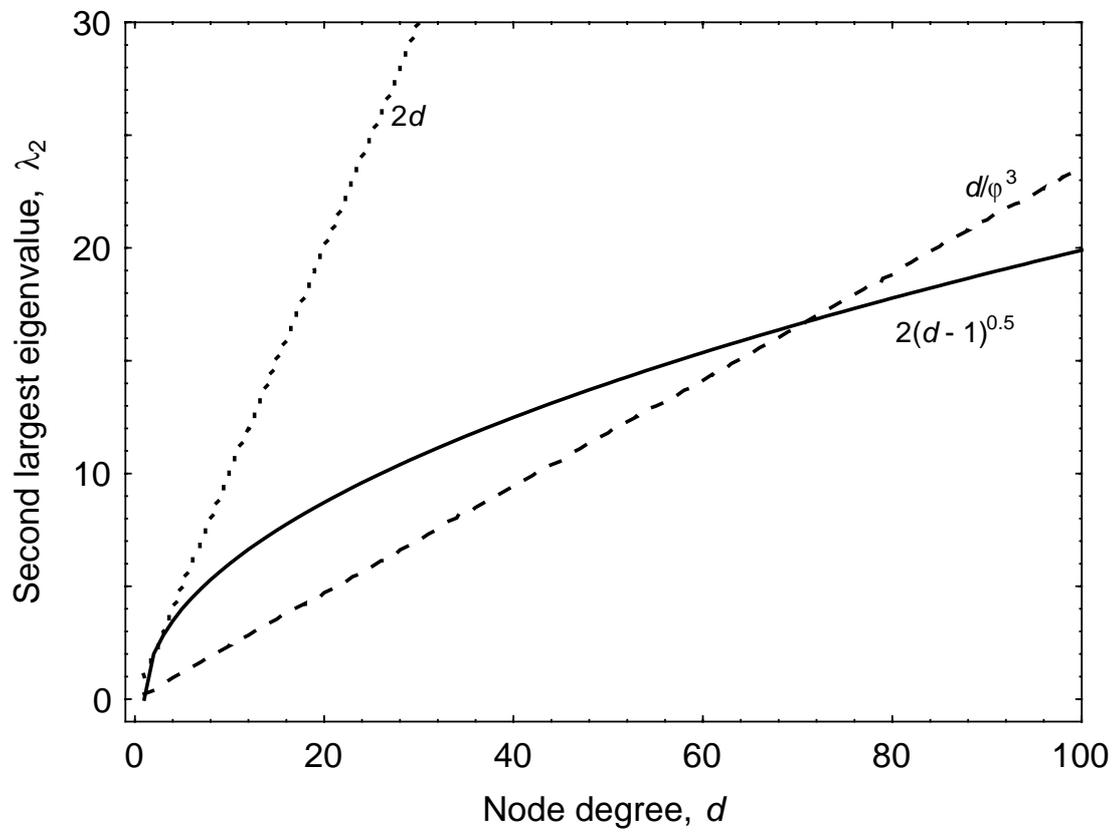

Fig. 4

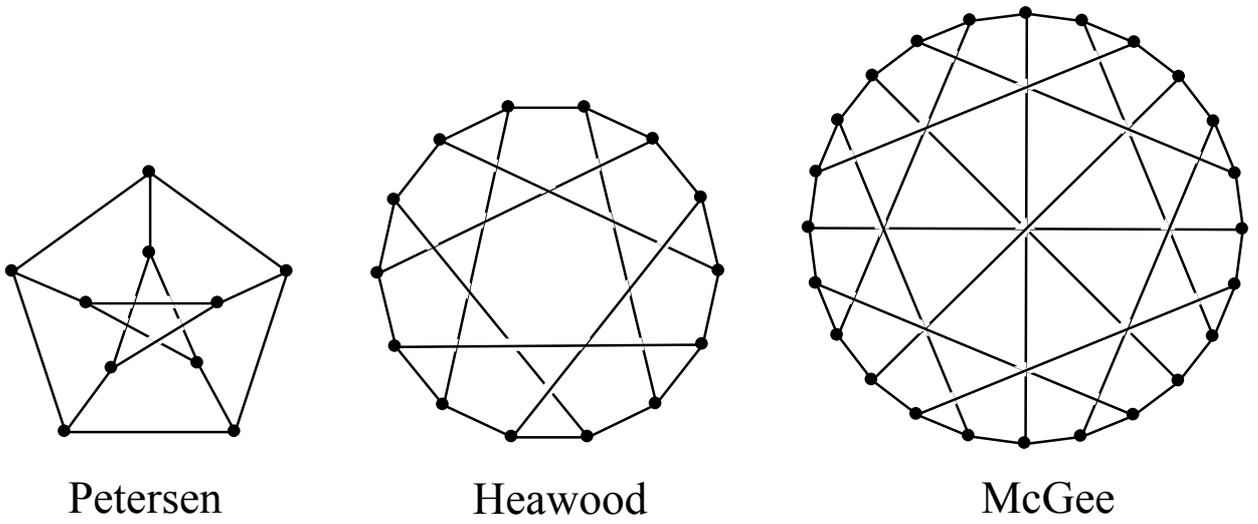

Petersen    Heawood    McGee

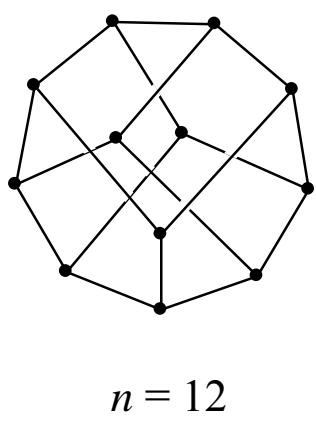  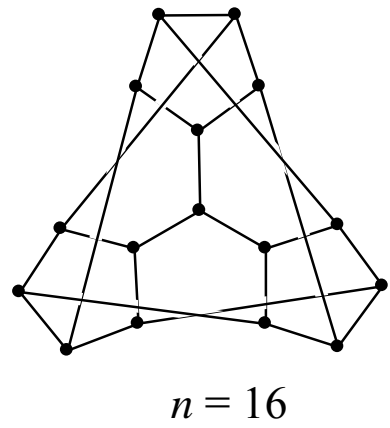

$n = 12$    $n = 16$

Donetti *et al*. entangled graphs

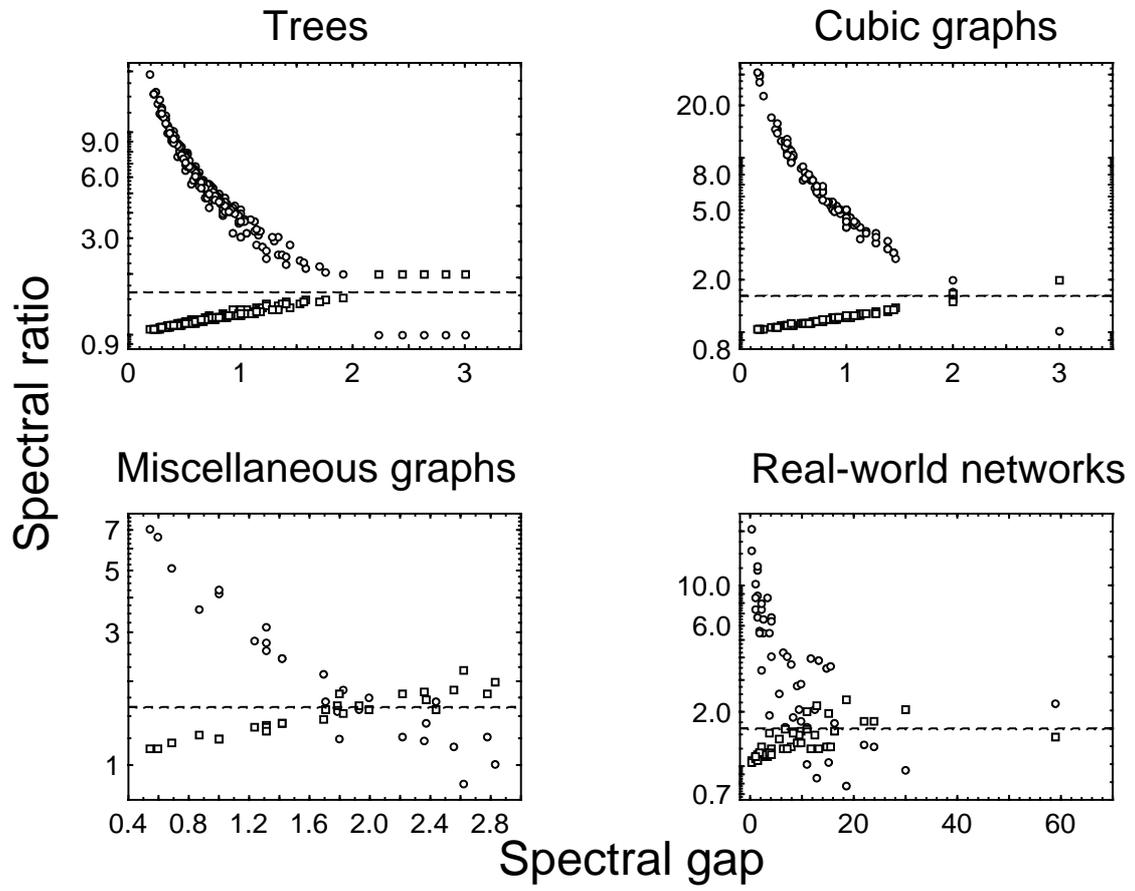

Fig. 6

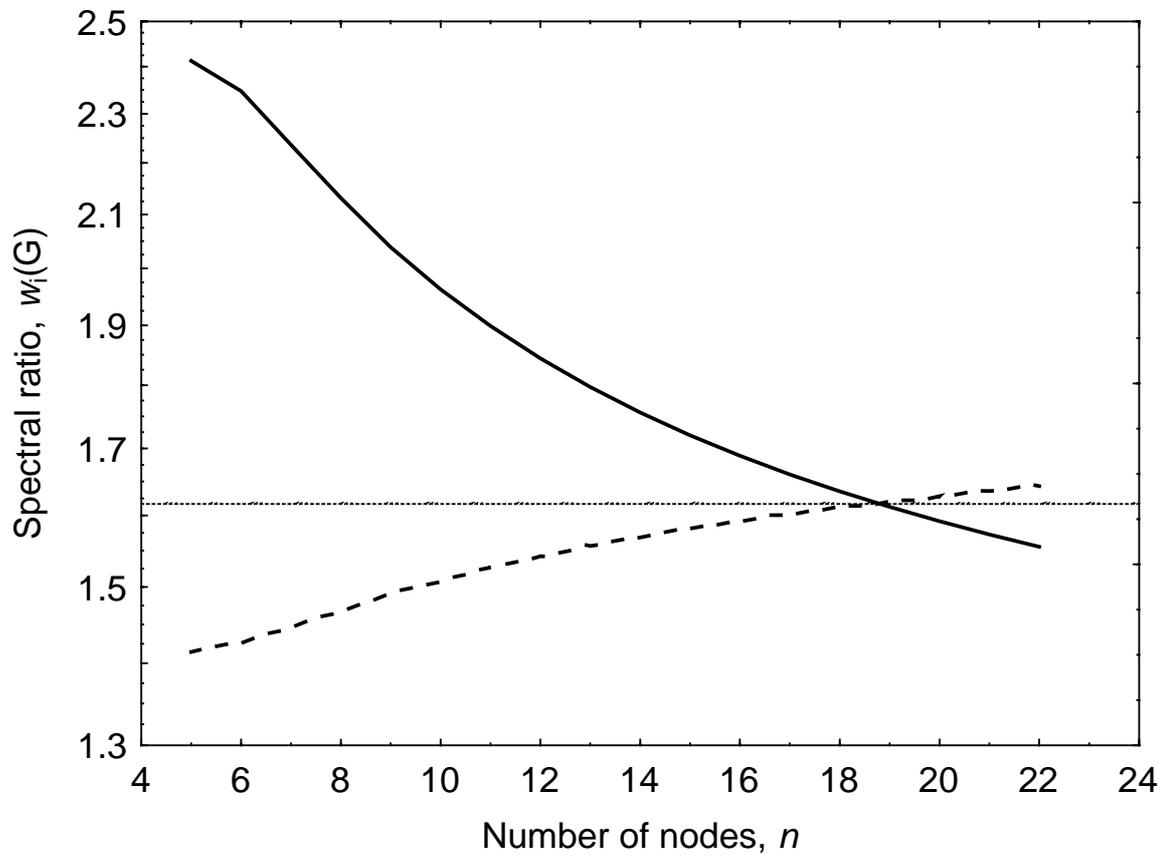

Fig. 7

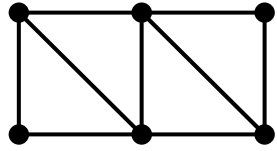  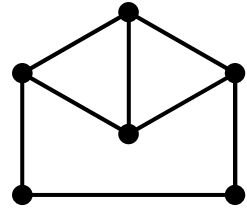

a   b

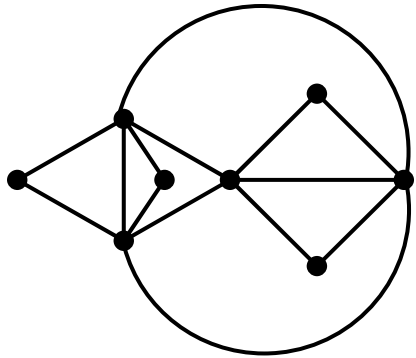   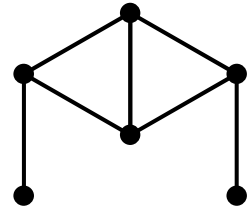

c   d

Fig. 8

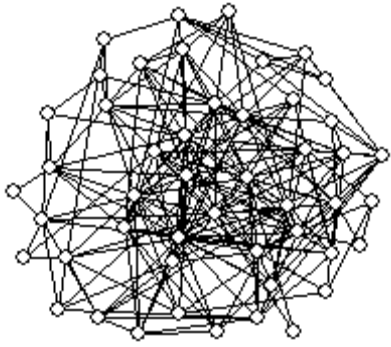 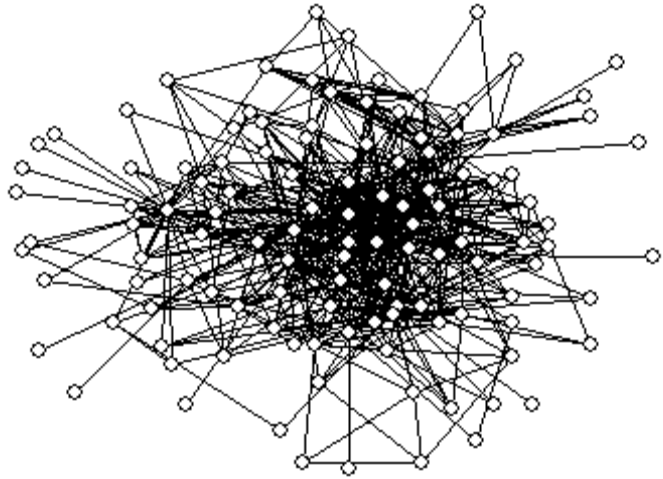

**"St. Marks" food web**     **"Centrality" citation network**